\documentclass{elsart}
\usepackage{graphicx}
\usepackage{color}

\begin{document}

\newcommand{\beq}{\begin{equation}}
\newcommand{\eeq}{\end{equation}}
\newcommand{\beqn}{\begin{equation}}
\newcommand{\eeqn}{\end{equation}}
\newcommand{\mrm}{\mathrm}
\newcommand{\mbf}{\mathbf}
\newcommand{\ol}{\overline}
\newcommand{\fr}{\frac}
\newcommand{\kT}{k_\mrm{B}T}
\newcommand{\kB}{k_\mrm{B}}

\begin{frontmatter}

\title{Changes in the electrical transport of
ZnO under visible light}

\author{S. Dusari*, J. Barzola-Quiquia, P. Esquinazi.}
\corauth[cor]{srujana@physik.uni-leipzig.de}
 \address{Division of Superconductivity and Magnetism, Institut für Experimentelle Physik II,
Universit\"at Leipzig, Linnestr. 5, D-04103, Germany.}

\author{S. P. Heluani}
 \address{Laboratorio de F\'{i}sica del S\'{o}lido,  Dpto. de
F\'{i}sica, Facultad de Ciencias Exactas y Tecnolog\'{i}a,
Universidad Nacional de Tucum\'{a}n, Argentina.}

\begin{abstract}
Complex impedance spectroscopy data in the frequency range 16Hz $ <
f < 3$MHz at room temperature were acquired on pure ZnO single
crystal and thin film. The measured impedance of the ZnO samples
shows large changes with time after exposure to or covering them from
visible light. At fixed times Cole-Cole-diagrams indicate the
presence of a single relaxation process. A simple analysis of the
impedance data allows us to obtain two main relaxation times. The
behavior for both, ZnO crystal and thin film, is similar but the thin
film shows shorter relaxation times. The analysis indicates the
existence of two different photo-active defects with activation
energies between $\sim 0.8$ eV and $\sim 1.1$ eV.

\end{abstract}

\end{frontmatter}

\section{Introduction}

Zinc Oxide (ZnO) is emerging as a material of interest for a variety
of electronic applications. It is used in various areas like opto
electronics in the blue/ultra-violet range, pharmaceutical industry,
UV detectors and as a bio-compatible material
\cite{Klingshirn:PRB07,Ozgur:JAP05}. This semiconductor has gained
substantial interest in the community in part because of its wide
direct band gap ($\sim 3.37$ eV at room temperature
\cite{Ozgur:JAP05}). Due its large band gap and small spin orbit
coupling, a long spin life time is expected. Therefore, ZnO is a
potential candidate for room temperature spintronic applications
\cite{Dietl:SCI00,Karpina:CRT04,Carcia:APL03}.
 Advantages associated with the large band gap include higher breakdown voltages, ability to sustain large
electric fields, lower noise generation, and high temperature and high-power operation \cite{Ozgur:JAP05}. Apart from these characteristics it is chemically stable, easy to prepare and nontoxic.
Most of the doping materials that are used with ZnO are also readily available.

Although the DC conductivity in pure and doped ZnO samples has received
considerable attention in the last years, there have been only a few studies on their AC
transport properties. Even though that optical absorption and emission properties of
ZnO were already studied recently \cite{Carcia:APL03}, the effect of visible light on the electrical
conductivity has received minor attention. Taking into account that a promising
application of ZnO is its use in transparent electronic devices, such as in the fabrication
of transparent thin-film transistors \cite{Lorenz:Sping07}, it appears advisable to study the electrical
impedance of ZnO samples as a function of time after exposure to or covering from
visible light. Our results show that for undoped ZnO samples the influence of the
photoconductivity to the electrical transport is large and cannot be rejected. Therefore,
care should be taken with time-dependent effects on the related properties due to the
influence of visible light.

We would like to note that our results obtained only at room temperature provide
a guide of the expected changes in the electrical transport in a broad frequency range
due to the effect of visible light. This work does not intend to provide a detailed studied
on the nature of the defects in pure ZnO crystals and films, a unsolved issue nowadays,
but to demonstrate the capability of the used experimental method and the large
observed effects in the frequency dependence of the conductivity due to visible light. To
the best of our knowledge such a frequency study and the influence of visible light was
not yet reported in the literature for ZnO. Studies as a function of the light wavelength
and temperature are out of the scope of our work and can be done in the future using a
similar experimental method as described here.

\section{Experimental Details.}

We measured ZnO single crystals obtained from the company Crystec
GmbH grown by hydrothermal method and with dimensions $5\times 5
\times 1$~mm$^{3}$. The thin film of 110 nm thickness was prepared by
pulsed laser deposition (PLD) on a r-plane sapphire substrate using a
substrate temperature of 400~C and oxygen pressure of 0.002~mbar. The
grown parameters correspond to the same as those used to produce
high-quality thin films by PLD method reported in Ref.
\cite{Lorenz:Sping07}. The single phase of the used thin film sample
was verified by using X-ray diffraction, which spectra is shown in
Fig. 1. The single ZnO peak corresponds to the (1 1 0) plane of the
wurtzite structure.

\begin{figure}
\begin{center}
\includegraphics[scale=1]{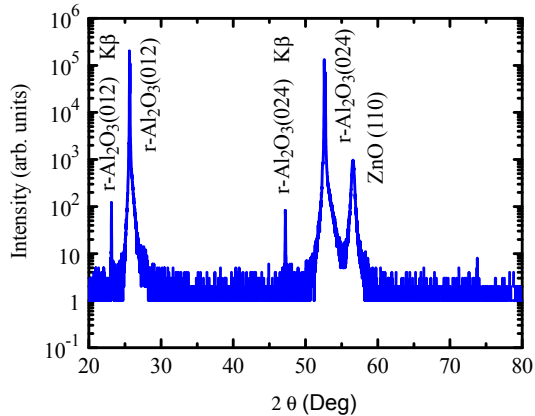}
 \caption{XRD spectra for the ZnO thin film grown on r-plane sapphire substrate.}
\end{center}
\end{figure}

In order to attain high-performance ZnO-based optical and electrical
devices it is essential to achieve Ohmic contacts that have both low
resistance and are thermally stable and reliable \cite{Ozgur:JAP05}.
It is well known that parasitic resistance, in the form of contact
resistance, is one of the major obstacles in realizing long-lifetime
operation of optical and electrical devices. The major loss of device
performance is often caused by high resistance metal-semiconductor
contacts through contact failures having potential barriers and/or
thermal stress. Four-point probe configuration was used for all
electrical measurements in a symmetrical device using gold wires
($25\mu$m diameter) fixed with pure indium on the ZnO surface, as
shown schematically in Fig. 2(a). To check for a linear, ohmic
behavior of the sample-contact system we have measured the I-V
characteristic curves using a Agilent semiconductor analyzer (4155C).
Stepwise steady state voltage biased measurements were carried out at
room temperature. The obtained I-V curves show a linear and symmetric
relationship in a wide range of voltages, see Figs. 2(b) and (c),
ensuring that our contacts show Ohmic behavior.

\begin{figure}
\begin{center}
\includegraphics[scale=1]{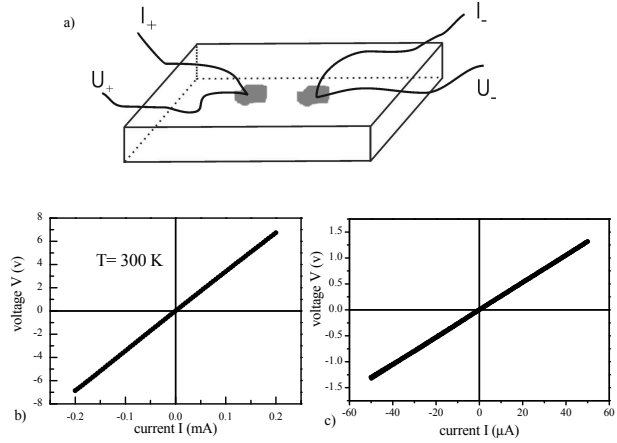}
 \caption{a) Schema of the contacts location on the sample
 surface.b) Current-Voltage characteristics for
 the ZnO single crystal at room temperature.c) Current-Voltage
 characteristics for the ZnO thin film at room temperature.}
\end{center}
\end{figure}

Impedance measurements were carried out using  Agilent 4294A
impedance analyzer (in the frequency range of 40~Hz to 3~MHz) and a
16~Hz AC resistance LR700 (Linear Research) bridge at room
temperature, in dark as well as under visible light (typical
luminance $\sim 400~$lx) as a function of time. We measured the
impedance, phase, resistance, and reactance using alternating signals
up to 500 mV amplitude. The impedance spectra were obtained as
follows: the samples were exposed to visible light for nearly 5 hours
until the signal was in a steady state situation (no significant time
change) and after a first measurement they were covered. Then,
measurements in dark were taken at successive times. All measurements
in this work were performed at room temperature in a acclimatized
environment ($297~$K$ \pm 1~$K).

\section{Results}

Short time dependent resistance measurements for one of the single crystals are presented in Fig. 3, where we demonstrate the reversibility of the effect of exposure the crystal with light below the ZnO band gap energy and then covering the sample. Other
crystals from the same batch show the same behavior. It is remarkable to see that in the first 75 seconds after covering the sample a clear increase in the resistance of the order of 15\% is observed, see Fig. 3.

\begin{figure}
\begin{center}
\includegraphics[scale=1]{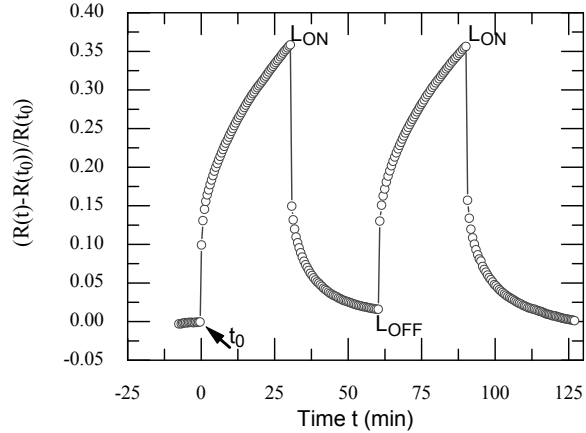}
 \caption{Relative change of the resistance vs. time for the ZnO single crystal after
exposure to ($L_{on}$) and covering from ($L_{off}$) visible light. The ac current frequency in this
measurement was 16 Hz.}
\end{center}
\end{figure}

We present below the results of the complex impedance as $Z=Z^{'}+ i
Z^{''}$, where $Z^{'}$ (resistance) and $Z^{''}$ (reactance) are the
real and imaginary parts of the impedance. The real part of the
impedance $Z^{'}$ versus frequency for the single crystal is shown in
Fig. 4(a) where the black curve represents the frequency response of
the sample during frequencies. Under light and in dark the real part
of the impedance shows a frequency independent region extending from
DC to $\sim 20~$kHz followed by a decrease at higher illumination and
in the dark, see Fig. 4(a). The decrease in the real part of the
impedance with frequency is in agreement with reported results in the
literature \cite{Thangavel:ML07}. Figure 4(b) shows the variation of
the imaginary part of impedance $Z^{''}$ (reactance) with frequency
at similar times as in Fig. 4(a). The spectra are characterized by
the appearance of a minimum, which shifts to lower frequencies with
time. Such behavior indicates the presence of relaxation processes in
the system as has been observed in, e.g. Ti-doped LSMO
\cite{Rahmouni:JMMM07}.

 \begin{figure}
\begin{center}
\includegraphics[scale=1]{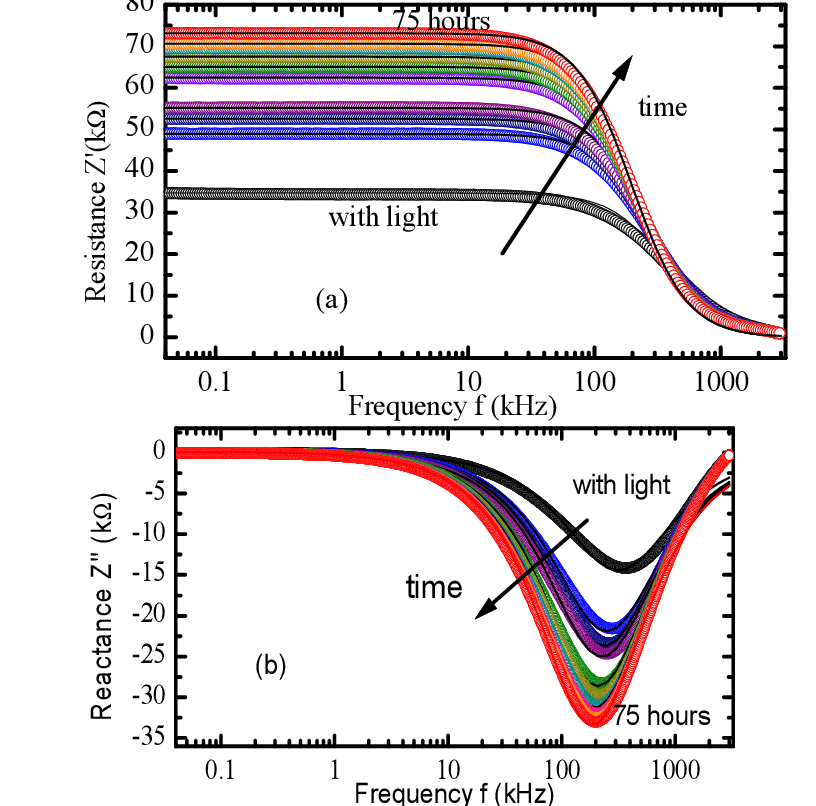}
 \caption{(a)Variation of the real part of the impedance $Z^{'}$ of the ZnO single crystal as a function of frequency at a steady state under light exposure and at different times after covering the sample (dark state). (b) The same but for the
imaginary part of the impedance $Z^{'}$. The curves are obtained in light, and in dark after 2 h, 4h, 6h, 17h, 24h, 36h, 48h, 63h, 75h. Solid lines across the points are the calculated curves using Eqs.(1) and (2).}
\end{center}
\end{figure}

Figure 5 shows the variation of the reactance $Z^{''}$ as a function of the resistance $Z^{'}$, i.e. Cole-Cole plot, at room temperature at different times after covering the sample. At each time the spectrum appears as a semicircle diagram indicating that a single relaxation mechanism is present \cite{Barsoukov:WI05}. Note the independence of the semicircle diagram from time. Such pattern tells us about the electrical processes occurring within the sample and their correlation with the sample microstructure. In order to analyze the experimental results, a simple equivalent circuit was used to represent the measured system, i.e. a parallel RC circuit connected with an induction L in series.
 According to this model the expression of the real ($Z^{'}$) and imaginary ($Z^{''}$) components are:

\begin{equation}
 Z^{'}=\frac{R}{(1+R^{2} \omega^{2} C^{2})}  \\
\end{equation}
\begin{equation}
 Z^{''}=\frac{R^{2} \omega C}{(1+R^{2} \omega^{2} C^{2})}+ L
 \omega\,.
\end{equation}

After fitting the experimental curves using the above model, see Fig.4, we have obtained
the capacitance for each measurement at a given time. Figure 6 shows the calculated
capacitance as a function of time t after covering the sample. We observe that there is
an exponential decay of the capacitance with time. This time dependence can be fitted
with a combination of two exponential functions: $C(t)=C(0)+a \ e^{-t/\tau_{1}} + b \ e^{-t/\tau_{2}}$,
with $\tau_{1}$ and $\tau_{2}$ two independent relaxation times and a,b free parameters. From the fit to
the data, see Fig. 6, we obtain $\tau_{1}\approx 0.56$ hours and $\tau_{2} \approx 22$ hours.

\begin{figure}
\begin{center}
\includegraphics[scale=1]{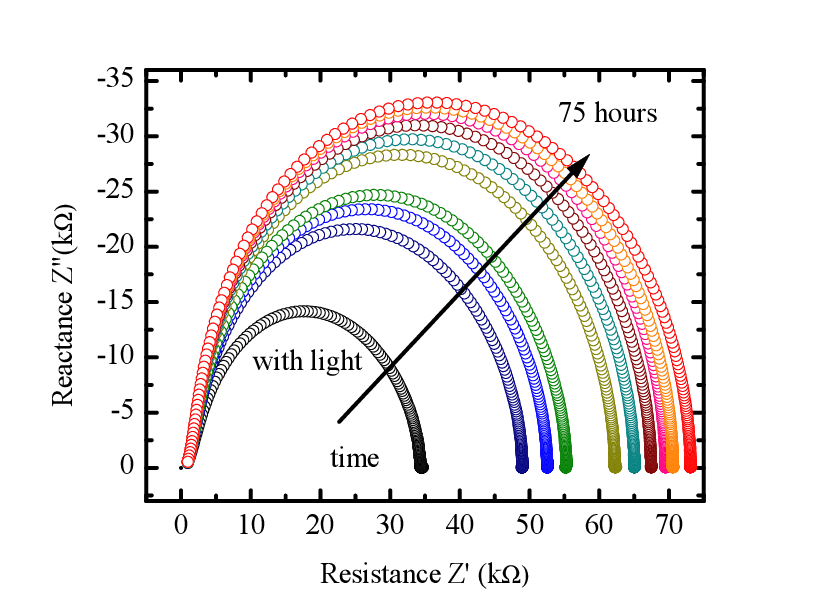}
 \caption{Cole-Cole plot of the AC impedance diagram of ZnO single crystal in light and dark conditions at different times after exposure it to visible light. The times are similar to those in Fig.4.}
\end{center}
\end{figure}

In order to check for the existence of similar effects in thin films we measured the
impedance spectra for the ZnO thin film also. Figures 7(a) and 7(b) show the variation of
the real and imaginary part of the impedance with frequency when the sample is
exposure to and after covering it from visible light. The behavior of these plots is similar
to that of the single crystal as well as the Cole-Cole plots, see Fig. 8. Figure 9 shows the
capacitance versus time obtained from the fits shown in Fig. 7. Also for the thin film the
capacitance decreases exponentially after covering the sample from visible light. The
behavior can be fitted with two relaxation times of values 0.1hs and 11hs. These
relaxation times are smaller as well as the relative change of the transport properties
compared to the values obtained for the single crystal. This result indicates that the
structural quality of the samples play a role. From one side a larger defect concentration
in the ZnO film compared to the single crystal produces larger scattering then smaller
 electron mobility. On the other hand the carrier density in the thin film is larger than in
the single crystal due to the lattice disorder. These changes due to the structural quality
affect the photoconductivity. Similar dependence of the photoconductivity with the
microstructure was obtained in TiO2 films \cite{Comedi:JPCM07}.

\begin{figure}
\begin{center}
\includegraphics[scale=1]{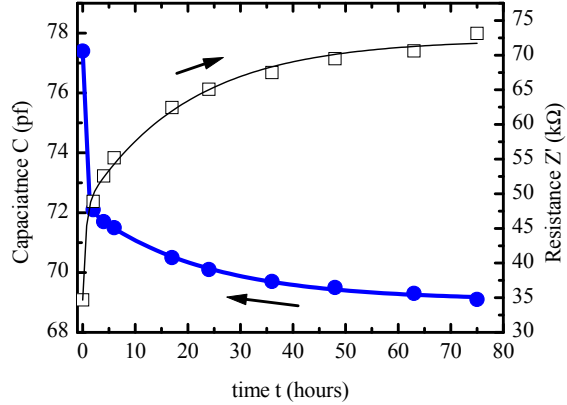}
 \caption{Time dependence of the obtained capacitance (\textcolor{blue}{$\bullet$}) and resistance ($\Box $) for the
ZnO single crystal. The continuous line through the data was obtained with a second
order exponential function with the relaxation times of 22h and 0.56h.}
\end{center}
\end{figure}

\begin{figure}
\begin{center}
\includegraphics[scale=1]{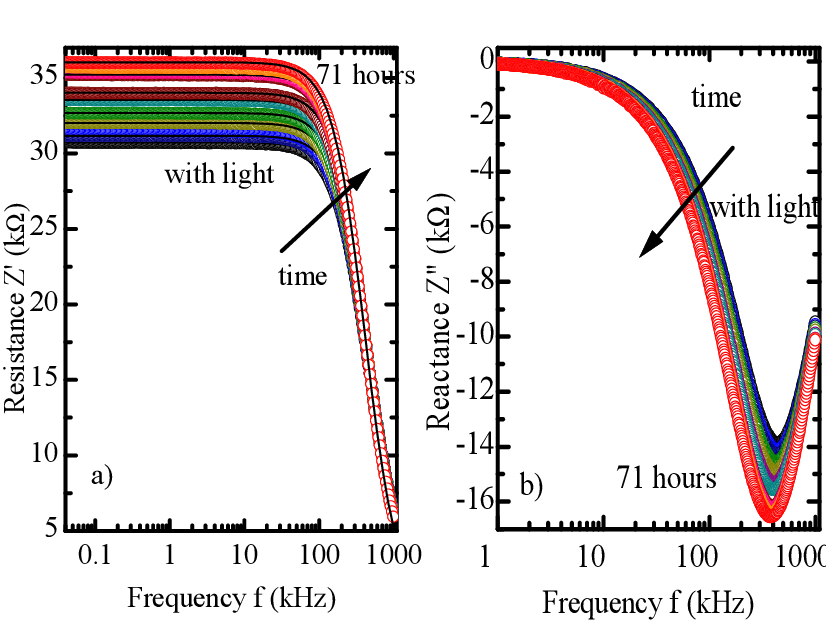}
 \caption{(a) Variation of the real part of the impedance $Z^{'}$ for the ZnO thin
film as a function of frequency at a steady state under light exposure and at different
times after covering the sample (dark state).Solid lines are the calculated curves
using Eqs.(1) and (2). (b) The same but for the imaginary part of the impedance $Z^{'}$.
The curves are obtained in light and in dark after 15 min, 30 min, 1h, 2 h, 4h, 6h, 17h,
23h, 29h, 47h, 71h.}
\end{center}
\end{figure}

\begin{figure}
\begin{center}
\includegraphics[scale=1]{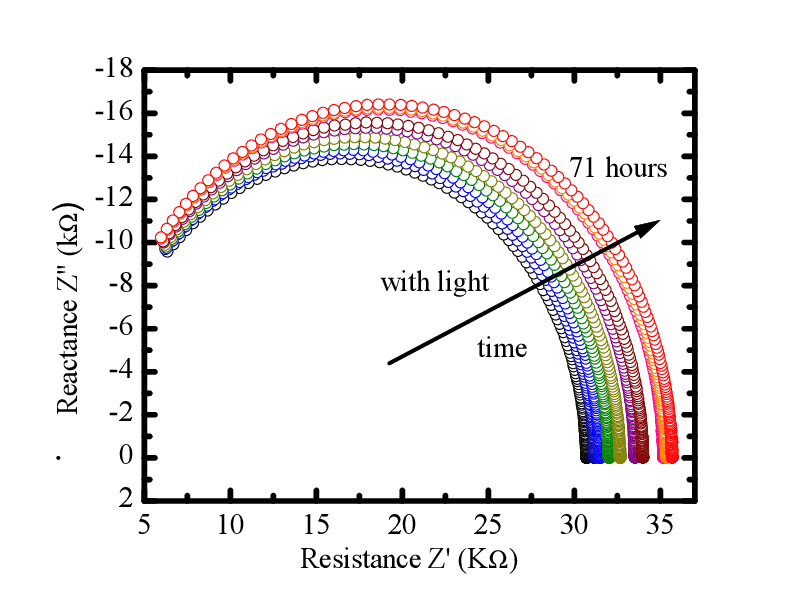}
 \caption{Cole-Cole plot of the AC impedance diagram of ZnO thin film in light and dark conditions at different times after exposure it to visible light. The times are similar to those in Fig. 7.}
\end{center}
\end{figure}

\begin{figure}
\begin{center}
\includegraphics[scale=1]{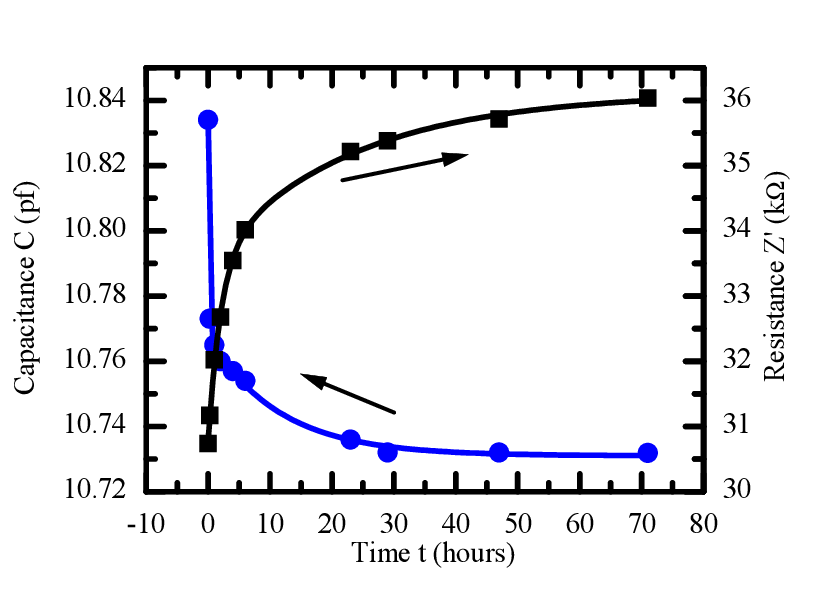}
 \caption{Variation of the obtained capacitance and measured resistance versus time for the ZnO thin film. The continuous line through the capacitance data is obtained using an exponential function second order with the relaxation times of 0.1h and 11h. The continuous line through the resistance data is a guide to the eye.}
\end{center}
\end{figure}

\section{Discussion}

As pointed out in the introduction a detailed study of the defects
that are active in the photoconductivity process of ZnO crystals and
films is out of the scope of this work. Neither the nature of the
defects nor their behavior under irradiation in this compound is
still understood nor does our fixed temperature allow us a detailed
study of the activation processes. Therefore we give some general
remarks on our results. Let us start with the effects produced by the
irradiation of visible light to a semiconductor like ZnO. The time
dependent response depends on the corresponding relaxation time of
the photon absorption or emission processes
\cite{Grundmann:Spring06}, processes that are influenced by intrinsic
and extrinsic lattice defects. During sub-gap illumination excess
carriers are optically excited from localized band gap states and
quickly thermalize in the extend states. They will be many times
trapped in the localized states and send back to the conduction
extended states. After a short time, the shallower state will be in
thermal distribution. Eventually, several of them could be trapped in
deeper levels and remains frozen. A distribution of shallower traps
could be responsible for the short time photoconductivity whereas the
slower photoconductivity response could be due to capture of holes in
deep trap levels. Which might be the origin of those defect levels?
Usually ZnO has n-type character even in the absence of intentional
doping. Native defects such as oxygen vacancies or zinc interstitial
are often assumed to be the origin of this, but the subject remains
controversial \cite{Look:PRL99}. Recent work indicates, for example,
the existence of a huge amount of hydrogen ($\sim 0.3 \%$) in ZnO
single crystals \cite{Brauer:PRB09}. These "impurities" can influence
the electronic properties of ZnO as has been observed several years
ago \cite{Heiland:SSC69}. On the other hand, in relation to the
photoconductivity, it was already suggested that point defects
behaving as hole or electron traps, exhibit slow electron-hole
recombination \cite{Moazzami:SST06} compared to other defects.

The behavior obtained for both samples indicates that there should be
at least two kinds of photoactive defects in the ZnO samples that are
responsible for the behavior of ZnO under visible light. Some of the
active defects could be related to the vacancy-hydrogen complexes
recently found in ZnO crystals \cite{Brauer:PRB09}. Also cation and
anion vacancies which are double acceptors and donors in ZnO may
contribute  \cite{Meyer:PSS04}. Assuming that the processes that
contribute to the time dependence we observed are thermally
activated, i.e. $\tau_{1} = \tau_{0} e^{E_a/k_{B}T}$) with $E_{a}$
the activation energy and $\tau_{0}^{-1}$
 an attempt frequency, we estimate activation energies $E_{a}$ = 0.85 eV – 0.97eV (0.81 eV –
0.93 eV) and 0.95 eV - 1.07 eV (0.93 eV –- 1.05 eV) for the short –-
large relaxation times and assuming as attempt frequencies $10^{11}
Hz$ and $10^{13} Hz$ for the single crystal (thin film),
respectively. These energies are comparable to those found in the
literature \cite{Moazzami:SST06}. Future experimental studies should
try to measure the photoconductivity at certain wavelengths and at
different temperatures for a complete characterization of the
photoactive processes and the defects that play a role in this
phenomenon.

\section{Conclusions}
The impedance spectra of ZnO single crystals and a thin film have been obtained after
exposure the samples to visible light and after covering them at room temperature and
in a broad frequency range. From the measured impedance parameters we obtained
two main relaxation times, a short one of less than 1 hour and a long one above 10
hours, after covering the sample. We found a clear difference in the relaxation times
between the single crystal and thin film that should be related to the structural quality of
the samples. The observed large visible-light dependence in the electrical impedance in
a broad frequency range of both investigated samples can be a limitation for real
applications.

\ack This work is supported by the DAAD/PROALAR under PKZ D/08/11707
and by the SFB 762 "Funktionalit\"at Oxidischer Grenzfl\"achen". S.D.
is supported by BuildMoNa and J.BQ. by the DFG under DFG ES 86/16-1.
We thank D. Spoddig, G. Bridoux, M. Ziese and A. Setzer for the
support with the measurements and M. Lorenz, H. Hochmuth and M.
Grundmann for providing us with the thin film sample.

\end{document}